\documentclass[pra,floatfix,twocolumn,superscriptaddress,showpacs,preprintnumbers,longbibliography,nofootinbib]{revtex4-1}

\usepackage[utf8]{inputenc} 
\usepackage{bibunits} 
\usepackage{amssymb,amsfonts,amsmath}
\usepackage{float} 
\usepackage{amsthm}
\usepackage{enumitem}
\usepackage[mathscr]{euscript}
\usepackage{bbold}
\usepackage[normalem]{ulem}
\usepackage{hyperref} 
\hypersetup{colorlinks=true, urlcolor=blue, citecolor=blue,linkcolor=blue}
\usepackage[all]{hypcap} 

\usepackage{graphicx}
\usepackage{dcolumn}
\usepackage{bm}
\usepackage{psfrag}
\usepackage{epsfig}
\usepackage{multirow}
\usepackage{color}
\usepackage{bbm}
\usepackage[FIGTOPCAP,raggedright,nooneline]{subfigure}
\usepackage{verbatim}
\usepackage{upgreek} 
\usepackage{physics}
\usepackage{tikz} 

\newcommand{\T}[1]{\text{#1}}

\newcommand{\ignore}[1]{}

\newcommand{\eq}{Eq.\,}
\newcommand{\eqs}{Eqs.\,}
\newcommand{\fig}{Fig.\,}

\newcommand{\cf} {cf.~}

\newcommand{\ie} {i.e.~}
\newcommand{\eg} {e.g.~}
\newcommand{\rref} {Ref.\,}

\begin{document}

	\title{Quantum correlations beyond entanglement in a classical-channel model of gravity}

	\author{Federico Roccati}
	\affiliation{Department of Physics and Materials Science, University of Luxembourg, L-1511 Luxembourg}
	
	\author{Benedetto Militello}
	\affiliation{Universit$\grave{a}$ degli Studi di Palermo, Dipartimento di Fisica e Chimica -- Emilio Segr$\grave{e}$, via Archirafi 36, I-90123 Palermo, Italy}
	\affiliation{INFN Sezione di Catania, via Santa Sofia 64, I-95123 Catania, Italy}
	
	\author{Emilio Fiordilino}
	\affiliation{Universit$\grave{a}$ degli Studi di Palermo, Dipartimento di Fisica e Chimica -- Emilio Segr$\grave{e}$, via Archirafi 36, I-90123 Palermo, Italy}
	
	\author{Rosario Iaria}
	\affiliation{Universit$\grave{a}$ degli Studi di Palermo, Dipartimento di Fisica e Chimica -- Emilio Segr$\grave{e}$, via Archirafi 36, I-90123 Palermo, Italy}

	\author{Luciano Burderi}
	\affiliation{Dipartimento di Fisica, Universit$\grave{a}$ degli Studi di Cagliari, SP Monserrato-Sestu, KM 0.7, 09042 Monserrato, Italy}
	
	\author{Tiziana Di Salvo}
	\affiliation{Universit$\grave{a}$ degli Studi di Palermo, Dipartimento di Fisica e Chimica -- Emilio Segr$\grave{e}$, via Archirafi 36, I-90123 Palermo, Italy}
	
	\author{Francesco Ciccarello}
	
	\affiliation{NEST, Istituto Nanoscienze-CNR, Piazza S. Silvestro 12, 56127 Pisa, Italy}
	\affiliation{Universit$\grave{a}$ degli Studi di Palermo, Dipartimento di Fisica e Chimica -- Emilio Segr$\grave{e}$, via Archirafi 36, I-90123 Palermo, Italy}
	
	\date{\today}
	
	\begin{abstract}
		A direct quantization of the Newtonian interaction between two masses is known to establish entanglement, which if detected would witness the quantum nature of the gravitational field.
		Gravitational interaction is yet compatible also with gravitational decoherence models relying on classical channels, hence unable to create entanglement. Here, we show in paradigmatic cases that, despite the absence of entanglement, a classical-channel model of gravity can still establish quantum correlations in the form of quantum discord between two masses. This is demonstrated for the Kafri-Taylor-Milburn (KTM) model and a recently proposed dissipative extension of this. In both cases, starting from an uncorrelated state, a significant amount of discord is generally created. This eventually decays in the KTM model, while it converges to a small stationary value in its dissipative extension. We also find that initial local squeezing on the state of the masses can significanlty enhance the generated discord.
		
	\end{abstract}
	
	\maketitle
	
	\section{Introduction}
	
	Among the four fundamental forces, gravity is the only one whose quantum nature was never demonstrated. Currently, there is growing confidence that the hindrances to testing the quantumness of the gravitional field due to Planck-scale limits could be overcome through table-top experiments \cite{Taylor2019tabletop}. Some of these, in particular, propose that detecting entanglement between two masses would witness the quantum nature of the gravitational field mediating their mutual interaction \cite{BosePRL2017,MarlettoPRL2017,carlesso2019testing}. 
	
	To date, as no phenomena are yet known where quantum mechanics and gravity coexist, the possibility that gravity could be just classical cannot be ruled out. If so, however, as the quantum nature of matter is well-established one should yet conceive a hybrid scenario where a classical channel mediates gravitational interactions between intrinsically quantum masses. Several models of this kind have been proposed \cite{kafriAQ2013, kafriNJP2014,tilloyPRD2017,altamirano2017unitarity,BassiKTM21,DissKTM}. In these models, typically, the conjectured classical channel gives rise to the Newtonian potential, yet causing at the same time decoherence affecting the quantum masses \cite{bassiCQG2017}. Decoherence plays against the entanglement that would arise from the Newtonian potential, the net result being that no entanglement can be generated in such models in agreement with their classical nature \cite{kafriAQ2013, kafriNJP2014,tilloyPRD2017,altamirano2017unitarity,BassiKTM21,DissKTM}. More specifically, this follows from the fact that they rely only on local operations and classical communication (LOCC), namely operations unable to create entanglement (this being a distinctive property of entanglement itself) \cite{nielsen2010}.

	Notwithstanding the above, classical channel models of gravity could be still compatible with establishment of quantum correlations (QCs). Indeed, entanglement is not the most general form in which correlations of a non-classical nature can manifest. This has been known since the early 2000s, when it was introduced a general quantifier of QCs usually going under the name of {\it quantum discord} or simply ``discord''~\cite{ollivierPRL2001,hendersonJPAMG2001}. Notably, while any entangled state has non-vanishing discord, the converse does not hold: there are states which -- although fully separable (non-entangled) -- still feature non-local correlations incompatible with classical physics. Remarkably, these  zero-entanglement QCs can be harnessed as a resource for a number of quantum information processing tasks, which was confirmed in a number of experiments~\cite{modiRMP2012,bera2017quantum}.
	Remarkably, unlike entanglement, discord {\it can} be created through LOCC. For instance, a local dissipative channel can turn a classically-correlated state into one with non-zero discord (but still disentangled) \cite{ciccarelloPRA2012a,streltsovPRL2011}. 
	
	With the above motivations, this work addresses the question as to  whether or not quantum correlations according to this extended paradigm can be generated in a classical channel model of gravity and, if so, whether they are stable or eventually decay at large times. 
	We carry out this task in the case study of the Kafri Taylor Milburn (KTM) model \cite{kafriNJP2014} and its recently proposed dissipative version \cite{DissKTM}. This allows to consider a relatively simple system made out of a pair of quantum harmonic oscillators for which effective techniques were developed to compute quantum discord~\cite{giordaPRL2010,adessoPRL2010} (calculation of discord is generally quite challenging~\cite{modiRMP2012,bera2017quantum}). We will in particular show that, although entanglement never shows up, starting from a fully uncorrelated state QCs are indeed created during the dynamics in a significant amount (compared to the total correlations). Such generated discord eventually undergoes a slow decay in the case of the KTM model, while it converges to a finite, although small, stationary value for its dissipative extension in \rref\cite{DissKTM}.
	\\
	\\
	
	The paper is organized as follows. In Section \ref{potential}, we introduce the system (two suspended masses) and shortly review the standard linearization and quantization of their Newtonian potential. Next, in Section \ref{sec-KTM} we shortly review the KTM model discussing in particular the related master equation for the two-mass system. We work out the ensuing differential equation for the covariance matrix, which is the essential quantity needed for the calculation of discord in the case of Gaussian states. Section \ref{sec-comp} recalls the definition and calculation of quantum discord. In Section \ref{sec-disc1}, we study the dynamics of QCs in the KTM model for coherent and squeezed initial states of the masses, investigating the dependence on the amount of squeezing in the latter case. 
	In Section \ref{sec-disc2}, we study creation of QCs in the dissipative version of the KTM model recently introduced in \rref\cite{DissKTM} (which is first reviewed).
	Finally, we present our conclusions in Section \ref{sec-concl}.

	\section{Quantized Newtonian potential}\label{potential}
	
	\begin{figure}
		\centering
		\includegraphics[width=4.cm]{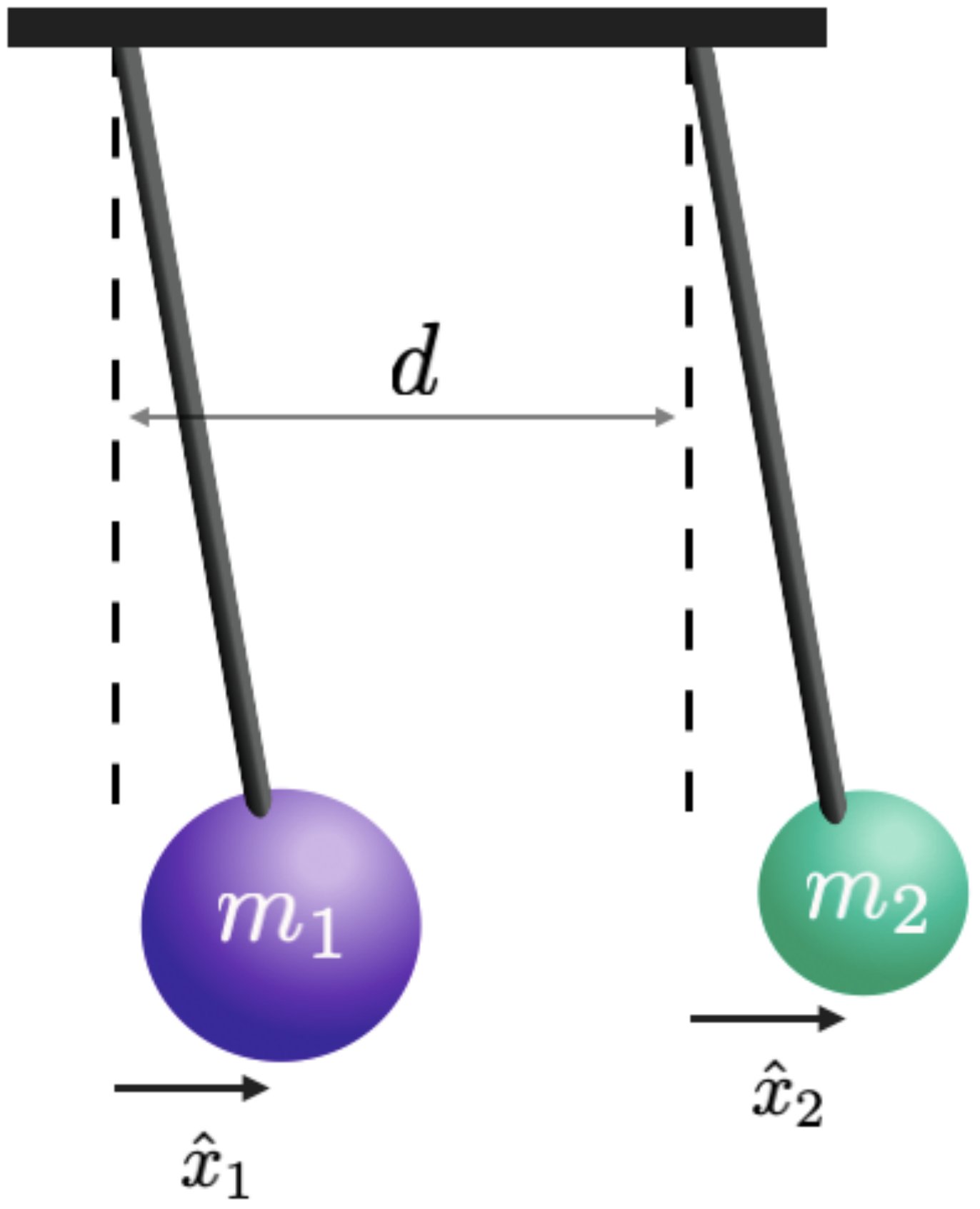}
		\caption{System: two suspended masses (pendula) $m_1$ and $m_2$, where $x_j$ is the displacement of the $j$th mass from the respective equlibrium position and $d$ the distance between the equilibrium positions.} 
		\label{setup}
	\end{figure}
	
	Consider two suspended masses $m_1$ and $m_2$ (see \fig\ref{setup}) subject to mutual Newtonian attraction.  In the usual regime of small oscillations, the masses are effectively modeled as a pair of independent quantum harmonic oscillators so that the Hamiltonian reads
	\begin{equation}
		\hat H=\hat H_0-G\,\frac{m_1m_2}{\sqrt{\hat x_{21}^2}}\,,\label{H1}
	\end{equation}
	with $\hat H_0$ the free Hamiltonian given by
	\begin{equation}\label{H0}
		\hat H_0 = \sum_{j=1}^2 \left(\frac{\hat p_j^2}{2m_j} + \frac{1}{2} m_j\omega_j^2\,\hat x_j^2\right)\,
	\end{equation}
	where $G$ the gravitational constant while $\hat x_{21}=\hat x_2-\hat x_1$.
	
	In the standard regime of small displacements from the equilibrium positions and redefining $\hat x_j$ as the displacement of mass $m_j$ from the equilibrium position (i.e., the one for $G\neq 0$), Hamiltonian \eqref{H1} is well-approximated by
	\begin{equation}\label{H}
		\hat H= \hat H'_0+ \hat V
	\end{equation} 
	where $\hat H'_0$ is obtained from $\hat H_0$ [\cf\eq\eqref{H0}] by replacing $ \omega_j^2$ with $\omega_j^2 - K/m_j$, while $\hat V$ embodies the (linearized) Netwon's interaction Hamiltonian
	\begin{equation}
		\hat V=K\,\hat x_1\hat x_2\,\label{V}
	\end{equation}
	with
	\begin{equation}\label{Kdef}
		K=2G\, \frac{m_1m_2}{d^3}\,,
	\end{equation}
	where $d$ is the distance between the equilibrium positions of the two masses\footnote{\eq\eqref{H} is obtained by expanding the Newtonian potential to the 2nd order in the mass displacements as	\begin{equation}\label{Vdef}
			\hat V\simeq  -G\,\frac{m_1m_2}{d} \left(1+\frac{\hat x_{21}}{d}-\frac{\hat x_{21}^2}{d^2}\right)\,
		\end{equation}
		and eliminating next the linear terms through the aforementioned redefinition of each mass coordinate and frequency.}.
	
	Potential \eqref{V} results from the mere quantization of the {\it static} Newtonian potential irrespective of what channel or charge carriers mediate the gravitational attraction. 
	In this model, the two masses jointly embody a closed system (no decoherence). Accordingly, their dynamics is unitary and thus governed by the Liouville-von Neumann equation (equivalent to the Schr\"odinger equation) 
	\begin{eqnarray}\label{VN}
		\dot\rho & = & -i[\hat H ,\rho] 
	\end{eqnarray}
	with $\rho$ the joint density operator of the two masses and $\hat H$ given by \eqref{H}. Here and throughout the paper we set $\hbar=1$.
	\\
	\\	
	As generally expected for quantum systems subject to a direct mutual interaction and isolated from an external environment, the Newtonian potential $\hat V$ causes establishment of entanglement between the two masses when these start in a fully uncorrelated state, which was shown in \rref \cite{carlesso2019testing}.

	\section{KTM model}\label{sec-KTM}
	
	The KTM model \cite{kafriNJP2014a} assumes a specific, fully {\it classical}, channel mediating the gravitational interaction. This gives rise to the Newtonian potential $\hat V$ which is however accompanied by additional decoherence affecting the two-mass system. Specifically, the classical channel consists of local measurements plus feedback as briefly sketched next (see \eg\rref\cite{Altamirano_2018} for a more detailed description). The position of $m_1$ is first instantaneously measured, the outcome being $x_1$. A local field described by a Hamiltonian term $\propto\!x_1 \hat x_2$ is then applied on mass 2 for an infinitesimal time (note that here only $\hat x_2$ is an operator). Analogous operations with swapped roles of systems 1 and 2 are next carried out and the whole process iterated over and over during the entire time evolution. Note that such a dynamics is necessarily stochastic since measurements in quantum mechanics are probabilistic. The average (so called ``unconditional") dynamics, however, is deterministic and demonstrably described by the master equation (ME) \cite{kafriNJP2014a}
	\begin{equation}\label{uncond3}
		\dot\rho = -i[\hat H,\rho]  
		+\sum_{j=1}^2 \left(\lambda+ \frac{K^2}{4\lambda}\right)\mathscr D[\hat x_j]\rho
	\end{equation}
	with $\hat H$ and $K$ the same as \eqref{H} and \eqref{Kdef}, respectively, and where we set $\mathscr D[\hat O]\rho=\hat O\rho\,\hat O^\dagger -\frac{1}{2}(\hat O^\dagger\hat O\rho+\rho\,\hat O^\dagger\hat O)$. Here, $\lambda$ measures the characteristic rate of the measurement-feedback operation\footnote{This rate is generally dependent on the susbsystem ($m_1$ or $m_2$). Here, we  assumed it to be independent of the subsystem for the sake of simplicity.}. The terms containing $\lambda$ (jointly called dissipator) describe decoherence affecting the two masses, which makes the system effectively open and its dynamics non-unitary. Notably, decoherence is minimized for $\lambda=K/2$, a value which we will set throughout the remainder.
	Without dissipator, ME \eqref{uncond3} would reduce to \eqref{VN}, showing that the classical channel gives rise to the canonical gravitational attraction yet introducing at the same time ineliminable decoherence (observe that the dissipator is non-zero for any value of $\lambda$).
	
	We point out that the dissipator in \eqref{uncond3} is the sum of two {\it local} dissipators: such ``local noise" (as is often referred to) is well-known to generally spoil entanglement. This counteracts the effect of the Hamiltonian term in ME \eqref{uncond3} which instead can create entanglement due to the Newtonian potential \eqref{V} as discussed in the previous section. The net result is that the dynamics described by ME \eqref{uncond3} is unable to create entanglement. Physically, this is due to the fully classical nature of the gravitational channel described above which relies solely on LOCC. In fact, therefore, the KTM model cannot generate entanglement by construction.

	\subsection{Rescaling}
	
	It is convenient to introduce rescaled positions and momenta as (recall  that $\hbar=1$) 
	\begin{equation}\label{rescale}
		\hat X_j =\sqrt{m\omega}\,\hat x_j\,,\,\,\,\hat P_j =\frac{\hat p_j}{\sqrt{m\omega}}\,\,.
	\end{equation}
	In terms of these dimensionless operators, master equation \eqref{uncond3} for $\lambda=K/2$ (minimal decoherence) can be arranged as 
	\begin{equation}\label{KTM}
		\dot\rho 
		= 
		-i[\hat H,\rho]  
		+
		\eta \omega\sum_{j=1}^2 \mathscr D[\hat X_j]\rho
	\end{equation}
	with
	\begin{equation}\label{Hform}
		\hat 	H=\omega
		\left[
		\frac{1}{2} \sum_{j=1}^2 \left(\hat P_j^2 + (1-\eta)\hat X_j^2 \right) 
		+
		\eta  \hat X_1\hat X_2
		\right]\,.
	\end{equation}
	Here, in line with other works \cite{krisnandaNQI2020} we introduced the dimensionless parameter
	\begin{equation}
		\eta=\frac{K}{m\omega^2}\label{eta}\,,
	\end{equation}
	namely the ratio between the gravitational coupling strength and the characteristic energy of the harmonic confinement of each mass. Thus $\eta$ measures the effective strength of the gravitational interaction. Since such interaction is typically very weak, $\eta$ should always be considered such that $\eta\ll 1$. In the remainder, we will frequently take advantage of this condition to make approximations.
	
	\subsection{Equation of motion for the covariance matrix}\label{sec-cov}
	
	We will consider throughout the two-mass system initially prepared in a {\it Gaussian state}, a class large enough to encompass many relevant states such as coherent, thermal and squeezed states. The form of the KTM master equation \eqref{uncond3}  is such that if the system starts in a Gaussian state then its state will remain Gaussian at any time \cite{ferraro2005}. 
	
	By definition, a Gaussian state of the two masses (quantum harmonic oscillators) is fully specified by the first and second moments $\langle \hat O_m \rangle$ and $\langle \hat O_m \hat O_n \rangle$ with $m,n=1,2,3,4$, where $\langle \hat A \rangle ={\rm Tr}\{\rho \hat A\}$ and $\hat O_1=\hat X_1$, $\hat O_2=\hat P_1$, $\hat O_3=\hat X_2$, $\hat O_4=\hat P_2$. For our purpose of calculating correlations (as will become clear later) it is sufficient to consider the covariance matrix $\sigma$ whose entries are defined as $\sigma_{mn} = \langle \hat O_m \hat   O_n+\hat O_n \hat O_m\rangle -2 \langle \hat O_m \rangle \langle \hat O_n \rangle$. 
	
	In the KTM model, master equation \eqref{uncond3} entails that $\sigma$ evolves in time according to the Lyapunov equation of motion~\cite{purkayastha2022lyapunov}
	\begin{equation}\label{eqforsigma_Y}
		\dot{\sigma} = Y\sigma +\sigma\, Y^{\rm T} + 4 D 
	\end{equation}
	with
	\begin{equation}
		Y
		=
		\left(
		\begin{array}{cccc}
			Y_{11}  &Y_{12} \\
			Y_{12}& Y_{11}  \\
		\end{array}
		\right)\,,
	\end{equation}
	where
	\begin{equation}
		Y_{11}
		=
		\omega 	\left(
		\begin{array}{cccc}
		0&1\\
			\eta {-}1  & 0
		\end{array}
		\right),\,Y_{12}
		=
		\left(
		\begin{array}{cccc}
			0  &  0\\
			-\eta \omega  & 0 
		\end{array}
		\right)\,\,,
	\end{equation}
	and with $D$ the diagonal matrix defined by\footnote{Note that all these matrices are symmetric under the exchange $1\leftrightarrow 2$ since so is master equation \eqref{KTM}. The same property for the same quantities also holds in the dissipative KTM model of Section \ref{sec-disc2} and, additionally, for any covariance matrix appearing throughout.}
		\begin{equation}
		D
		=
		\left(
		\begin{array}{cccc}
			D_{11}  &0 \\
		0& D_{11}  \\
		\end{array}
		\right)\,\,{\rm 	with}\,\,D_{11}
		= 	\left(
		\begin{array}{cccc}
			0&0\\
			0 & \tfrac{\eta \omega}{2}
		\end{array}
		\right)\,\,.
	\end{equation}	
	The solution of the linear equation of motion \eqref{eqforsigma_Y} reads
	\begin{equation}
		\sigma(t) = e^{Yt}\sigma_0 e^{Y^Tt} + 4\int_{0}^{t}\T{d}s\, e^{Ys}De^{Y^Ts}\label{sigmat}
	\end{equation}
	with $\sigma_0$ the covariance matrix at $t=0$ (the explicit analytical expression is cumbersome and thus not reported here).
	
	The knowledge of the covariance matrix at any time $t$ is enough to compute correlations of any kind (quantum or not) between the two masses.

	\section{Computation of quantum correlations}\label{sec-comp}

	The {\it total} amount of correlations between the two masses is measured by the mutual information ${\cal I}$~\cite{cover2006,nielsen2010}. This is defined in terms of the von Neumann entropy $S(\varrho)=-\T{Tr}(\varrho\log\varrho)$ with $\varrho$ a generic quantum state (represented by a density matrix). 
	If $\rho$ is the joint state of the two masses, mutual information $\mathcal I$ is the discrepancy between the sum of the local von Neumann entropies and the von Neumann entropy of the joint system according to
	\begin{equation}
	\mathcal I= S_{1}+S_{2}-S\,. \label{mutual}
	\end{equation}
Here, $S_{j}=-\T{Tr}(\rho_{j}\log\rho_{j})$ is the local entropy of mass $j$ [whose reduced state is given by $\rho_{1(2)}={\rm Tr}_{2(1)}\rho$], 
	while $S=-\T{Tr}(\rho\log\rho)$ is the joint entropy. A basic property of mutual information is that it vanishes if and only if $\rho=\rho_{1}\otimes\rho_{2}$, meaning that the two subsystems are fully uncorrelated. As such, mutual information captures all possible correlations regardless of their classical or quantum nature.
	
	In this work, our main focus is quantifying the amount of {\it quantum} correlations (QCs) which can be measured through the so called {\it quantum discord}~\cite{ollivierPRL2001,hendersonJPAMG2001,modiRMP2012}.
	This quantity expresses the discrepancy between ${\cal I}$ and another expression of the mutual information, ${\cal J}=S_1-S_{1|2}$ with $S_{1|2}$ the conditional entropy \cite{cover2006, nielsen2010} (amount of information to describe the state of subsystem 1 conditioned to a {\it local} measurement of 2), which provides the same result in a classical system, but different results in a quantum system.  Discord turns out to be evaluated as
	 		
	\begin{equation}
		\mathcal D=S_{1}-S+\underset{\hat M_{1k}}{\rm min}\,\,\sum_k p_k S(\rho_{2|k})\,.\label{d-def}
	\end{equation}
	
	Here, the minimization is over all possible quantum measurements $\{\hat M_{1k}\}$ performed on mass $j=1$. One such measurement with outcome $k$ 
	collapses the joint system onto $\rho_{2|k}=(\hat M_{1k} \rho\hat M_{1k} )/p_k$ with probability $p_k$\footnote{Note that discord is generally asymmetric, i.e., swapping indexes 1 and 2 in \eqref{d-def} generally changes the result. Yet, we will be interested throughout in dynamics and initial states which are invariant under the swap $1\leftrightarrow2$.}. 
	
	
	It is worth noting that this discrepancy arises from {\it non-local} correlations between the two subsytems, namely a genuinely non-classical feature: discord \eqref{d-def} is accordingly interpreted as a measure of the total amount of QCs.
	
	Most importantly, especially for our purposes here, while non-zero entanglement entails non-zero discord ${\cal D}$ the converse does not hold. Indeed, there exist {\it mixed} states which are separable (zero entanglement) but such that ${\cal D}\neq 0$\footnote{For instance, given two spin-1/2 particles each with basis $\{\ket{{\uparrow}}$, $\ket{{\downarrow}}\}$, the state $\rho=1/2 \ket{\uparrow}_{1}\!\bra{\uparrow}\otimes \ket{\uparrow}_{2}\!\bra{\uparrow}+1/2 \ket{+}_{1}\!\bra{+}\otimes \ket{\downarrow}_{2}\!\bra{\downarrow}$, where $\ket{+}=1/\sqrt{2}\,(\ket{\uparrow}+\ket{\downarrow})$, is separable but ${\cal D}\neq 0$. This is a consequence of the lack of distinguishability of states in quantum mechanics (in this case $\ket{\uparrow}$ and $\ket{+}$), which in turn follows from the superposition principle.}.
	
	For Gaussian states, which are the only ones entering our analysis (see Section \ref{sec-cov}), the minimization in~\eqref{d-def} can be restricted to Gaussian measurements~\cite{pirandolaPRL2014}, which produces a closed analytical expression of ${\mathcal D}$ (see Appendix~\ref{appendix}) as an explicit function of the $\sigma$'s entries~\cite{giordaPRL2010,adessoPRL2010} (so called Gaussian discord). 

	A property of Gaussian discord is that states such that $\mathcal D>1$ are entangled~\cite{adessoPRL2010}, which somehow reflects the aforementioned ability of discord to detect QCs more general than entanglement. A consequence of this is that for any classical channel (namely unable to create entanglement as in the KTM model) if ${\cal D}(0)<1$ then discord must remain below the threshold ${\cal D}=1$ at any time, that is ${\cal D}(t)<1$.

	\section{Creation of quantum correlations in the KTM model}\label{sec-disc1}
	
	Our aim is assessing the ability of the classical channel in the KTM model to create QCs and, if any, analyzing their temporal behaviour. Accordingly, we naturally focus throughout on initial product states, \ie of the form
	\begin{equation}
		\rho_0=\varrho_1\otimes\varrho_2 \label{prod}
	\end{equation}
	with $\varrho_j$ the initial state of mass $j$. State \eqref{prod} features no correlations of any kind between the two masses (thus mutual information ${\cal I}$ and of course ${\cal D}$ vanish).
	
	We will next consider two different types of initial states: coherent and squeezed.
	
	\subsection{Coherent states}
	
	When $\varrho_j$ is a coherent state it can be expressed as $\varrho_j=\ket{\alpha_j}\!\bra{\alpha_j}$ with $\ket{\alpha_j}=e^{\alpha_j \hat a_j^\dag-\alpha_j^*\hat a_j}\ket{0}_j$, where $\ket{0}_j$ the vacuum state and $\hat a_j$ the usual annihilation operator. 
	In this case, the covariance matrix corresponding to \eqref{prod} is simply given by 
	\begin{equation}
		\sigma_0=\T{diag}(1,1,1,1)\,,\label{diag}
	\end{equation}
	which is independent of the coherent-state amplitude $\alpha_j$.

\begin{figure}
		\centering
		\includegraphics[width=0.9\linewidth]{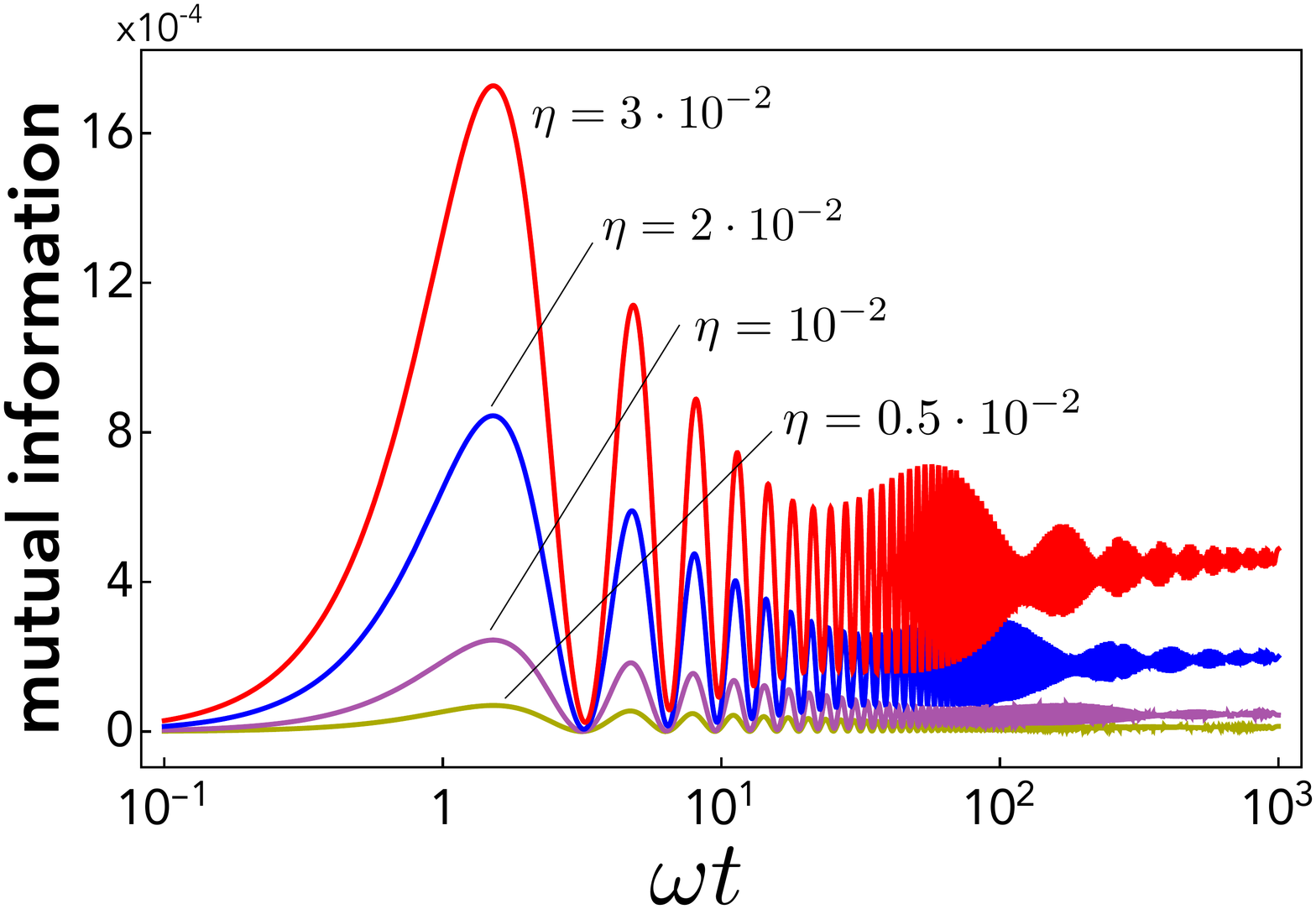}
		\label{mutual-coh}
		\caption{
			Semi-log plot of mutual information ${\cal I}$ against time (in units of $\omega^{-1}$) when each mass starts in a coherent state. Note that the growing frequency of damped oscillation is due to the logarithmic scale	.}
	\end{figure}

	To benchmark the dynamics and amount of quantum correlations to be analysed shortly, we first study the behaviour of mutual information ${\cal I}$ (measuring total correlations, either classical or quantum), which can be calculated straightforwardly from \eqs \eqref{mutual} and \eqref{diag}. The behaviour of ${\cal I}$ against time (rescaled in units of $\omega^{-1}$)  is reported in \fig\ref{mutual-coh} for growing values of $\eta$ [recall definition \eqref{eta}].
	We see that ${\cal I}$ first grows monotonically from zero until reaching its maximum value, then undergoes damped oscillations (exhbiting vanishing minima) and eventually asymptotically converges to a finite stationary value. Note that, as is reasonable, the maximum and asymptotic values grow with the effective gravitational interaction strength $\eta$.
	
\begin{figure}
		\centering
		\includegraphics[width=0.9\linewidth]{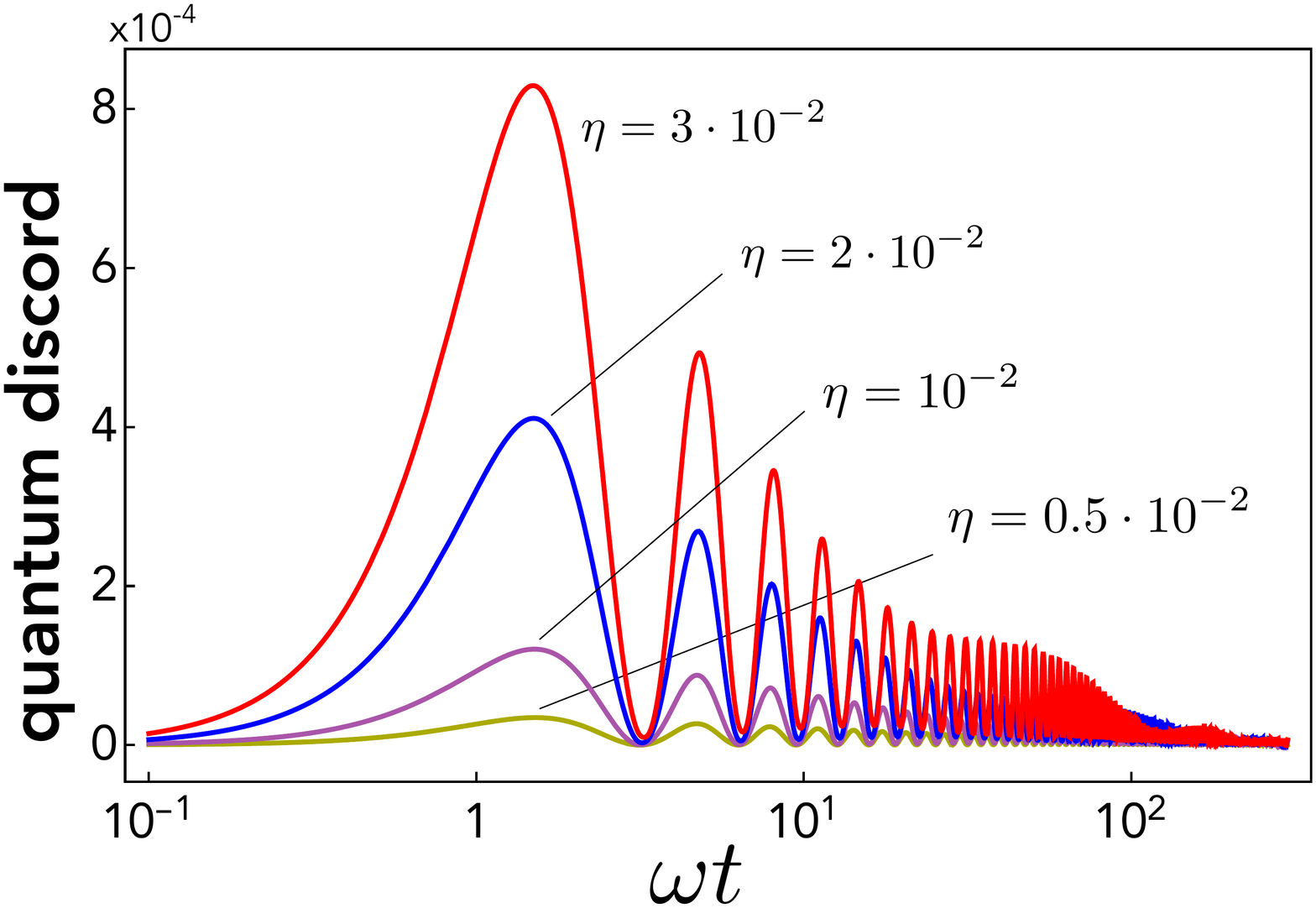}
		\label{discordKTM}
		\caption{
			Semi-log plot of quantum discord ${\cal D}$ against time (in units of $\omega^{-1}$) when each mass starts in a coherent state.}
	\end{figure}

	Let us now address the dynamics of quantum correlations. This is carried out by plugging $\sigma_0$ into \eqref{sigmat} and next working out the corresponding evolved Gaussian discord ${\cal D}(t)$, which yields an exact although cumbersome expression (not reported here). The time behaviour of ${\cal D}$ is shown in Fig.~\ref{discordKTM} for increasing coupling strengths $\eta$.
	Discord first grows monotonically until reaching a maximum and then undergoes damped oscillations with vanishing minima. 
	
	Thereby, quantum correlations indeed show up, thus confirming that they can be created by the classical gravitational channel. 
	The amount of quantum correlations established in the transient is significant, which can be seen \eg by noting that the first maximum of ${\cal D}$ is about half as that of mutual information [see \fig\ref{mutual-coh}].
	Overall, the temporal dynamics of discord is similar to mutual information except at long times. Indeed, the plots indicate that, within our numerical capabilities, quantum correlations are unstable, dropping off with a long tail for $t\rightarrow \infty$. 
	A rigorous proof that QCs vanish in the long-time limit is demanding due to the cumbersome expression of ${\cal D}(t)$ and the very long tail.
	Remarkably, however, this decay to zero can be shown analytically in the realistic regime of small $\eta$ (due to the weakness of gravitational interaction). Indeed, in this case, the long-time expression of discord can be worked out as
	\begin{eqnarray}\label{discordLargeTimesKTM}
		\mathcal{D}(\tau) \simeq  
		\frac{1}{2}
		\left[
		(1+\eta\tau) \log\frac{\eta\tau}{2+\eta\tau}
		+
		\log \left(1-(\eta\tau)^{-4}\right)\nonumber\right.\\\left.
		\qquad\qquad-
		\frac{(\eta\tau)^2}{1+\eta\tau} \log\frac{(\eta\tau)^2-\eta\tau-1}{(\eta\tau)^2+\eta\tau+1}
		\right]\,\,\,\,\,
	\end{eqnarray}
	with $\tau=\omega t$. This is easily shown to vanish in the limit $\tau\rightarrow \infty$.

To sum up, the above shows that quantum correlations are indeed created by the gravitational interaction in the transient and in a significant amount, but eventually (although slowly) fade away.
\begin{figure}
	\centering
	\includegraphics[width=0.9\linewidth]{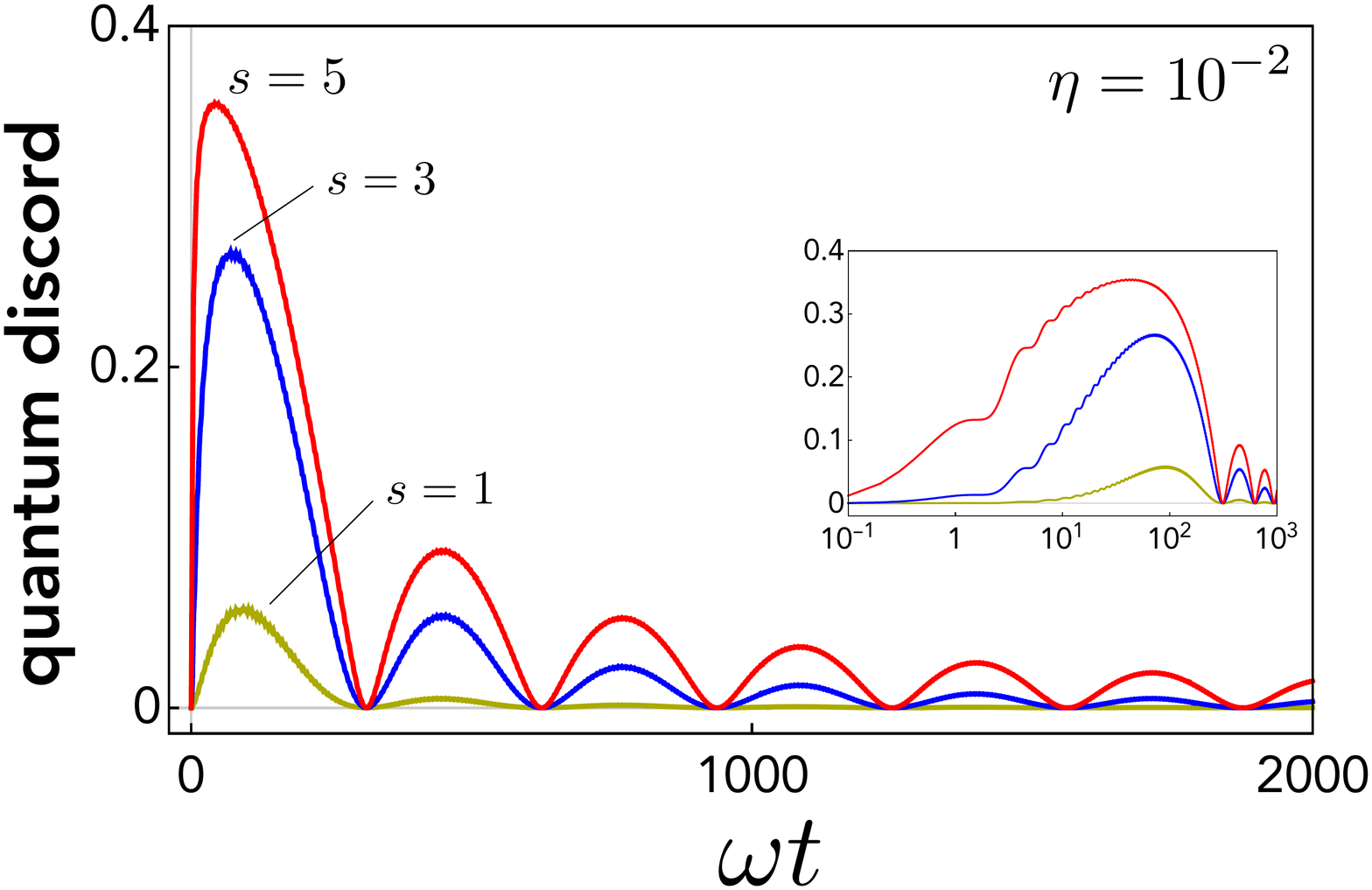}
	\label{discordKTMsqueez}
	\caption{
		Lin-lin plot of quantum discord against time (in units of $\omega^{-1}$) when each mass starts in a squeezed state $\ket{s_j}$ for different values of the squeezing parameter $s$. We set $\eta=10^{-2}$. The inset contains the lin-log version of the plot.}
\end{figure}

\subsection{Squeezed states}

Coherent states are the most classical states in that both the field quadratures $\hat x_i$ and $\hat p_j$ have minimum uncertainty. In contrast, we next take each $\varrho_j$ [\cf\eq\eqref{prod}] to be a squeezed state  ~\cite{gerry2005introductory} where the uncertainty on one of the two quadratures is ``squeezed'' (and the other one is consequently enhanced as imposed by the uncertaintly principle)\footnote{The non-classicality mentioned here lies in the reduced state of each oscillator (at $t=0$) and not in the correlations between the two oscillators [recall that initial state \eqref{prod} is fully uncorrelated].}.

A squeezed state of the $j$th mass reads $\eta_j=\dyad{s_j}$ with $\ket{s_j}=e^{[s_j (\hat a_j^2)^\dag-s_j^*\hat a_j^2]/2}\ket{0_j}$, where $s_j$ is the squeezing parameter. For simplicity, we assume the same squeezing on both the oscillators, i.e., $s_1=s_2=s$.  Accordingly, the covariance matrix corresponding to the initial state \eqref{prod} in this case can be worked out as
\begin{equation}\label{sigsq}
	\sigma_0
	=
	\left(
	\begin{array}{cc}
		S_2 & 0  \\
		0 & S_2  \\
	\end{array}
	\right)\,
\end{equation}
with
\begin{equation}
S_2
	=
	\left(
	\begin{array}{cc}
		\cosh s + \sinh s & 0  \\
		0 & \cosh s - \sinh s  \\
	\end{array}
	\right).
\end{equation}
For $s=0$ (no squeezing) $\sigma_0$ coincides with \eqref{diag} as in this limit the initial state of each mass reduces to the vacuum state (which can be seen as a zero-amplitude coherent state).

The calculation of discord \eqref{d-def} corresponding to \eqref{sigsq} yields again an involved expression, which is hard to arrange in a form as compact as Eq.~\eqref{discordLargeTimesKTM} even restricting to specific regimes.

In Fig.~\ref{discordKTMsqueez}, we set the representative value of coupling strength $\eta=10^{-2}$ and study the effect of squeezing (as measured by $s$) on generation of discord. A dynamics qualitatively similar to the case of coherent states [\cf\fig\ref{discordKTM}] occurs, which features an initial growth towards a maximum 
followed by damped oscillations characterized by a slow decay time.
A remarkable difference from coherent states [\cf\fig\ref{discordKTM}] yet stands out in that discord can reach far higher values during the transient, even orders of magnitudes larger (but still below the entanglement threshold ${\cal D}=1$, see Section \ref{sec-comp}). This is witnessed by the growth of maxima with $s$.

Interestingly, it was recently found in \rref\cite{krisnandaNQI2020} that squeezing also enhances entanglement generation if no decoherence is present, namely when the dynamics is unitary and described by the von Neumann equation \eqref{VN}. The present analysis thus indicates that discord (in place of entanglement) enjoys a similar property when the KTM dissipator is added to the ME [\cf\eq\eqref{KTM}].

\section{Dissipative KTM model}\label{sec-disc2}

As previousy discussed, the KTM model lacks a stationary state, namely there exists no solution of master equation \eqref{KTM} such that $\dot\rho=0$.  
This entails a temporal divergence of the energy of the masses. To overcome this pitfall, a dissipative version of the KTM model (henceforth referred to as DKTM model) was very recently proposed in \rref~\cite{DissKTM}. This model produces a master equation featuring the same Hamiltonian term as \eq\eqref{KTM} but a different dissipator in such a way that, differently from the KTM model, a stationary state exists (in particular preventing from energy divergence). As the original KTM model, this modified model still relies on LOCC operatios meaning that it still describes gravitational interactions mediated by a classical channel unable to create entanglement. 

In the following, after briefly reviewing the DKTM model's main features and associated master equation, we work out the ensuing equation of motion for the covariance matrix and then use it to investigate the dynamics of quantum correlations.

\subsection{ Review of the model}

Like the KTM model (\cf Section \ref{sec-KTM}), the DKTM model assumes that the gravitational interaction is mediated by a classical channel via local measurements and feedback. Yet, measurements of $\hat x_j$ (position of the $j$th mass) in the KTM model are now replaced by a measurements of the quadrature $\hat x_j + i \alpha \hat p_j$ with $\alpha$ a free parameter of the model that should be yet intended as small (the KTM model is retrieved for $\alpha=0$). 

The resulting master equation (see \rref~\cite{DissKTM} for details on the derivation) reads
\begin{eqnarray}
	\dot \rho 
	& = & 
	-i [\hat H+\delta \hat H,\rho] 
	+\eta \omega\sum_{j=1}^2 \mathcal D[\hat X_j]\rho\nonumber\\
	&  &
	- \tfrac{i}{2}\tilde \alpha \eta\omega\sum_{j} [\hat X_j,\{\hat P_j,\rho\}]	- \tfrac{1}{4} \tilde\alpha^2 \eta\omega  \sum_{j}  [\hat P_j,[\hat P_j,\rho]]
	\nonumber\\
	&  &	
	+  \tfrac{ 1 }{2}\tilde \alpha\eta\omega\sum_{j\neq j'}  [\hat X_j,[\hat P_{j'},\rho]]\,\label{jTMdissDimless}
\end{eqnarray}
with
\begin{equation}
\delta\hat H=\tfrac{1}{4}	\sum_{j} \tilde \alpha \eta\omega\, (\hat X_j \hat P_j+\hat P_j\hat X_j )\,,\,
\end{equation}
where $\hat X_j$ and $\hat P_j$ are rescaled positions and momenta [\cf\eq\eqref{rescale}], $\hat H$ and $\mathscr D$ are the same as in \eq\eqref{H} and where we rescaled $\alpha$ as $\tilde \alpha=m\omega \alpha$ (dimensionless).

It can be cheked that, setting $\alpha=0$, one recovers the master equation of the KTM model [\cf\eq\eqref{uncond3}].

\subsection{Quantum discord}

\begin{figure}
	\centering
	\includegraphics[width=1\linewidth]{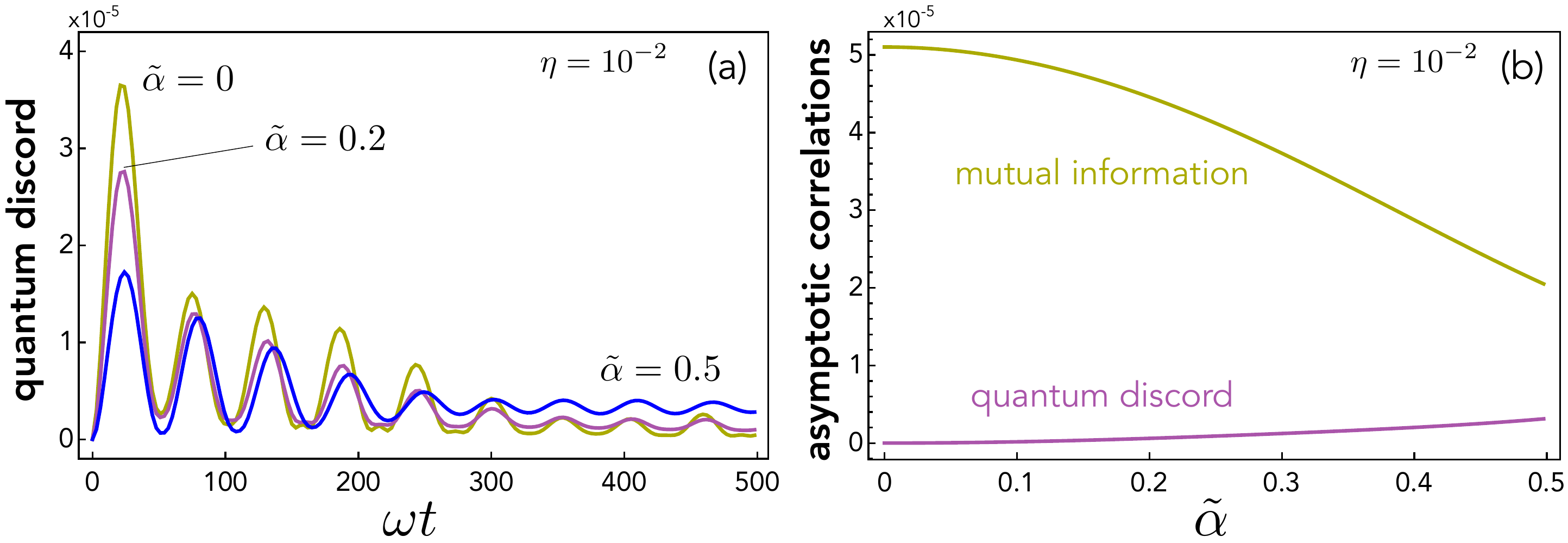}
	\label{disc-DKTM}
	\caption{Correlations in the dissipative KTM model. (a): Discord ${\cal D}$ against time (in units of $\omega^{-1}$) when each mass starts in a coherent state and for some representative values of parameter $\tilde{\alpha}$ (this must be small in fact by definition \cite{DissKTM}). The case $\tilde{\alpha}=0$ (standard KTM model) is also displayed for comparison. (b): Asymptotic discord (purple line) and mutual information (yellow) versus $\tilde{\alpha}$. In both panels we set $\eta=10^{-2}$.}
\end{figure}

The equation of motion for the covariance matrix $\sigma$ corresponding to master equation \eqref{jTMdissDimless}, which can be worked out analogously to the KTM model, is given by \eq\eqref{eqforsigma_Y} but with matrices $Y$ and $D$ now given by
\begin{equation}
	Y
	=
	\left(
	\begin{array}{cccc}
		Y_{11}  &Y_{12} \\
		Y_{12}& Y_{11}  \\
	\end{array}
	\right)\,
\end{equation}

with
\begin{equation}
	Y_{11}
	=
\omega 	\left(
	\begin{array}{cccc}
		\frac{1}{2} \tilde \alpha  \eta    &  1 \\
		(\eta -1)  & \,\,-\frac{3}{2} \tilde \alpha  \eta  
	\end{array}
	\right),\,Y_{12}
	=
	 	\left(
	\begin{array}{cccc}
		0  &  0\\
		-\eta \omega  & 0 
	\end{array}
	\right)\,\,,
\end{equation}
and
\begin{equation}
	D
	= 
	\left(
	\begin{array}{cccc}
		D_{11}  &D_{12} \\
		D_{12}& D_{11}  \\
	\end{array}
	\right)\,
\end{equation}
with
\begin{equation}
	D_{11}
	=
\tfrac{	\omega}{2} 	\left(
	\begin{array}{cccc}
	\frac{1}{2}\tilde \alpha ^2 \eta  &  0\\
	0 &\eta\omega  
	\end{array}
	\right),\,D_{12}
	=\tfrac{	\omega}{2} 
	\left(
	\begin{array}{cccc}
	0&	\frac{1}{2}\tilde \alpha ^2 \eta  \\
		\frac{1}{2}\tilde \alpha ^2 \eta  &  0
	\end{array}
	\right)\,\,.
\end{equation}
The typical time behaviour of quantum discord when each mass starts in a coherent state is reported in \fig\ref{disc-DKTM} for representative values of parameter $\tilde{\alpha}$, showing that discord is created even in the present model.
Somewhat similarly to the standard KTM model (\cf\fig\ref{discordKTM}) ${\cal D}$ grows from zero exhibiting secondary oscillations in the transient. In contrast to the KTM model, however, ${\cal D}$ rapidly saturates to a small stationary value that depens on $\tilde{\alpha}$. 

The occurrence of a steady value reflects the fact that the model is constructed so as to admit a stationary state. Indeed,
using standard methods of linear algebra, it can be shown that the matrix equation $\dot \sigma=0$ [\cf\eq\eqref{eqforsigma_Y}] admits only one solution $\sigma_\T{ss}$ (its analytical expression is too cumbersome to be reported here),
which is the covariance matrix of the unique steady state of master equation \eqref{jTMdissDimless}. The computed discord of $\sigma_\T{ss}$ using the parameters set in \fig\ref{disc-DKTM}(a) correctly matches the asymptotic value of ${\cal D}$ in each considered case.
This asymptotic value slightly grows with $\tilde\alpha$ as shown in \fig\ref{disc-DKTM}(b), where the asymptotic mutual information is also plotted for comparison. 
Overall, taking into account that $\tilde{\alpha}$ should be understood as a small parameter, we can conclude that the present model predicts that at large times classical correlations dominate over quantum ones somewhat in line with the KTM model.


\section{Conclusions}\label{sec-concl}
		
Whether the quantum nature of the gravitational field, if any, could be witnessed by detection of quantum correlations (QCs) between two massive objects is currently a hot theme. Within this general framework, some theories have been put forward which show that the gravitational interaction could be mediated by a classical channel relying on LOCC operations, hence unable to create entanglement. In this work, we asked whether such theories might still be compatible with establishment of QCs although not accompanied by entanglement (as measured by quantum discord). To this aim, we considered the Kafri Taylor Milburn (KTM) model and its associated master equation for a pair of quantum harmonic oscillators. Based on this, we predicted production of a significant amount of quantum discord, which yet eventually decays with time. Interestingly, an initial amount of local squeezing in the oscillators can greatly enhance the maximum value reached by discord in the transient, which is reminiscent of a similar property for entanglement recently shown in the case of quantum channels. Finally, we investigated the dynamics of QCs in a recently proposed dissipative extension of the KTM model, whose ensuing master equation admits a steady state (unlike the KTM model). Similarly to the KTM model, we showed that significant discord is created with a small fraction even surviving indefinitely.

We point out that, despite its usual detrimental action for entanglement generation, decoherence plays a key role in the creation of discord without entanglement. The reason is that discord can be non-zero for separable states only provided that these are mixed. In this respect, it is therefore essential for discord creation having a non-zero dissipator in the master equation in addition to the Newtonian interaction Hamiltonian.

These findings contribute to the general debate on the possible coexistence between classical gravity and quantum mechanics by showing that, even if the gravitational field were fully classical, it might still be able to establish non-classical correlations between quantum masses although of a non-entangled nature. It is worth noting that this further highlights the importance of detecting entanglement if the goal is demonstrating the quantum nature of the gravitational field, since other forms of quantum correlations can raise from classical gravitation.

{\it Acknowledgements.}---We acknowledge support through the FFR project ``Gravità e fisica quantistica"  funded by University of Palermo and DiFC.

		\appendix 
		
		\section{Calculation of quantum discord}\label{appendix}

		Due to the demanding minimization over all possible measurements, quantum discord as defined in \eq~\eqref{d-def} cannot be calculated explicitly in closed form for an arbitrary state, not even in the simplest case of a qubit (two-level system). However, for a pair of quantum harmonic oscillators which are in a two-mode Gaussian state, it can be shown \cite{adessoPRL2010} that it is not restrictive limiting the minimization to \textit{Gaussian} measurements, which yields a closed analytical formula. To give this, we first note that a generic $4\times 4$ covariance matrix has the form
		\begin{equation}
		\sigma
		=
		\left(
		\begin{array}{cc}
		A_1 & A_3  \\
		A_3^{\T{T}} & A_2  \\
		\end{array}
		\right)\,,
		\end{equation}
	where each $A_j$ is a $2\times 2$ matrix.
	    Defining now $I_j=\det A_j$ (for $j=1,2,3$), $I_4=\det \sigma$ and $\Delta=I_1+I_2+2 I_3$, $2\nu^2_\pm = \Delta\pm\sqrt{\Delta^2\pm4 I_4}$, discord \eqref{d-def} takes the closed form 
		\begin{equation}\label{discordeq}
		\mathcal D =
		f\left(I_2\right)
		-f\left(\nu_-\right)
		-f\left(\nu_+\right)
		+\delta
		\end{equation}
		where
		\begin{equation}
		\delta=
		\begin{cases}
		\frac{2I_3^2+(I_2-1)(I_4-I_1)+2|I_3|\sqrt{I_3^2+(I_2-1)(I_4-I_1)}}{(I_2-1)^2}  \\
		\qquad\text{if } (I_4-I_1I_2)^2<(I_2+1)(I_3+I_4)I_3^2 \\
		\frac{I_2I_1-I_3^2+I_4-\sqrt{I_3^4+(I_4-I_2I_1)^2-2I_3^2(I_4+I_2I_1)}}{I_2^2} \\
		\qquad\text{otherwise}
		\end{cases}
		\end{equation}
		Eq.~\eqref{discordeq} holds for measurements made on subsystem $2$ (for measurements on $1$ indexes 1 and 2 must be swapped).
		
%
%
%
%
%
%
%
%
%
%
%
%
		
		\bibliography{all,all_gravity}

\begin{thebibliography}{28}%
\makeatletter
\providecommand \@ifxundefined [1]{%
 \@ifx{#1\undefined}
}%
\providecommand \@ifnum [1]{%
 \ifnum #1\expandafter \@firstoftwo
 \else \expandafter \@secondoftwo
 \fi
}%
\providecommand \@ifx [1]{%
 \ifx #1\expandafter \@firstoftwo
 \else \expandafter \@secondoftwo
 \fi
}%
\providecommand \natexlab [1]{#1}%
\providecommand \enquote  [1]{``#1''}%
\providecommand \bibnamefont  [1]{#1}%
\providecommand \bibfnamefont [1]{#1}%
\providecommand \citenamefont [1]{#1}%
\providecommand \href@noop [0]{\@secondoftwo}%
\providecommand \href [0]{\begingroup \@sanitize@url \@href}%
\providecommand \@href[1]{\@@startlink{#1}\@@href}%
\providecommand \@@href[1]{\endgroup#1\@@endlink}%
\providecommand \@sanitize@url [0]{\catcode `\\12\catcode `\$12\catcode
  `\&12\catcode `\#12\catcode `\^12\catcode `\_12\catcode `\%12\relax}%
\providecommand \@@startlink[1]{}%
\providecommand \@@endlink[0]{}%
\providecommand \url  [0]{\begingroup\@sanitize@url \@url }%
\providecommand \@url [1]{\endgroup\@href {#1}{\urlprefix }}%
\providecommand \urlprefix  [0]{URL }%
\providecommand \Eprint [0]{\href }%
\providecommand \doibase [0]{http://dx.doi.org/}%
\providecommand \selectlanguage [0]{\@gobble}%
\providecommand \bibinfo  [0]{\@secondoftwo}%
\providecommand \bibfield  [0]{\@secondoftwo}%
\providecommand \translation [1]{[#1]}%
\providecommand \BibitemOpen [0]{}%
\providecommand \bibitemStop [0]{}%
\providecommand \bibitemNoStop [0]{.\EOS\space}%
\providecommand \EOS [0]{\spacefactor3000\relax}%
\providecommand \BibitemShut  [1]{\csname bibitem#1\endcsname}%
\let\auto@bib@innerbib\@empty
\bibitem [{\citenamefont {Carney}\ \emph {et~al.}(2019)\citenamefont {Carney},
  \citenamefont {Stamp},\ and\ \citenamefont {Taylor}}]{Taylor2019tabletop}%
  \BibitemOpen
  \bibfield  {author} {\bibinfo {author} {\bibfnamefont {D.}~\bibnamefont
  {Carney}}, \bibinfo {author} {\bibfnamefont {P.~C.~E.}\ \bibnamefont
  {Stamp}}, \ and\ \bibinfo {author} {\bibfnamefont {J.~M.}\ \bibnamefont
  {Taylor}},\ }\href {\doibase 10.1088/1361-6382/aaf9ca} {\bibfield  {journal}
  {\bibinfo  {journal} {Classical and Quantum Gravity}\ }\textbf {\bibinfo
  {volume} {36}},\ \bibinfo {pages} {034001} (\bibinfo {year}
  {2019})}\BibitemShut {NoStop}%
\bibitem [{\citenamefont {Bose}\ \emph {et~al.}(2017)\citenamefont {Bose},
  \citenamefont {Mazumdar}, \citenamefont {Morley}, \citenamefont {Ulbricht},
  \citenamefont {Toro\ifmmode~\check{s}\else \v{s}\fi{}}, \citenamefont
  {Paternostro}, \citenamefont {Geraci}, \citenamefont {Barker}, \citenamefont
  {Kim},\ and\ \citenamefont {Milburn}}]{BosePRL2017}%
  \BibitemOpen
  \bibfield  {author} {\bibinfo {author} {\bibfnamefont {S.}~\bibnamefont
  {Bose}}, \bibinfo {author} {\bibfnamefont {A.}~\bibnamefont {Mazumdar}},
  \bibinfo {author} {\bibfnamefont {G.~W.}\ \bibnamefont {Morley}}, \bibinfo
  {author} {\bibfnamefont {H.}~\bibnamefont {Ulbricht}}, \bibinfo {author}
  {\bibfnamefont {M.}~\bibnamefont {Toro\ifmmode~\check{s}\else \v{s}\fi{}}},
  \bibinfo {author} {\bibfnamefont {M.}~\bibnamefont {Paternostro}}, \bibinfo
  {author} {\bibfnamefont {A.~A.}\ \bibnamefont {Geraci}}, \bibinfo {author}
  {\bibfnamefont {P.~F.}\ \bibnamefont {Barker}}, \bibinfo {author}
  {\bibfnamefont {M.~S.}\ \bibnamefont {Kim}}, \ and\ \bibinfo {author}
  {\bibfnamefont {G.}~\bibnamefont {Milburn}},\ }\href {\doibase
  10.1103/PhysRevLett.119.240401} {\bibfield  {journal} {\bibinfo  {journal}
  {Phys. Rev. Lett.}\ }\textbf {\bibinfo {volume} {119}},\ \bibinfo {pages}
  {240401} (\bibinfo {year} {2017})}\BibitemShut {NoStop}%
\bibitem [{\citenamefont {Marletto}\ and\ \citenamefont
  {Vedral}(2017)}]{MarlettoPRL2017}%
  \BibitemOpen
  \bibfield  {author} {\bibinfo {author} {\bibfnamefont {C.}~\bibnamefont
  {Marletto}}\ and\ \bibinfo {author} {\bibfnamefont {V.}~\bibnamefont
  {Vedral}},\ }\href {\doibase 10.1103/PhysRevLett.119.240402} {\bibfield
  {journal} {\bibinfo  {journal} {Phys. Rev. Lett.}\ }\textbf {\bibinfo
  {volume} {119}},\ \bibinfo {pages} {240402} (\bibinfo {year}
  {2017})}\BibitemShut {NoStop}%
\bibitem [{\citenamefont {Carlesso}\ \emph {et~al.}(2019)\citenamefont
  {Carlesso}, \citenamefont {Bassi}, \citenamefont {Paternostro},\ and\
  \citenamefont {Ulbricht}}]{carlesso2019testing}%
  \BibitemOpen
  \bibfield  {author} {\bibinfo {author} {\bibfnamefont {M.}~\bibnamefont
  {Carlesso}}, \bibinfo {author} {\bibfnamefont {A.}~\bibnamefont {Bassi}},
  \bibinfo {author} {\bibfnamefont {M.}~\bibnamefont {Paternostro}}, \ and\
  \bibinfo {author} {\bibfnamefont {H.}~\bibnamefont {Ulbricht}},\ }\href
  {\doibase 10.1088/1367-2630/ab41c1} {\bibfield  {journal} {\bibinfo
  {journal} {New Journal of Physics}\ }\textbf {\bibinfo {volume} {21}},\
  \bibinfo {pages} {093052} (\bibinfo {year} {2019})}\BibitemShut {NoStop}%
\bibitem [{\citenamefont {Kafri}\ and\ \citenamefont
  {Taylor}(2013)}]{kafriAQ2013}%
  \BibitemOpen
  \bibfield  {author} {\bibinfo {author} {\bibfnamefont {D.}~\bibnamefont
  {Kafri}}\ and\ \bibinfo {author} {\bibfnamefont {J.~M.}\ \bibnamefont
  {Taylor}},\ }\href@noop {} {\bibfield  {journal} {\bibinfo  {journal}
  {arXiv:1311.4558 [quant-ph]}\ } (\bibinfo {year} {2013})},\ \Eprint
  {http://arxiv.org/abs/1311.4558} {arXiv:1311.4558 [quant-ph]} \BibitemShut
  {NoStop}%
\bibitem [{\citenamefont {Kafri}\ \emph
  {et~al.}(2014{\natexlab{a}})\citenamefont {Kafri}, \citenamefont {Taylor},\
  and\ \citenamefont {Milburn}}]{kafriNJP2014}%
  \BibitemOpen
  \bibfield  {author} {\bibinfo {author} {\bibfnamefont {D.}~\bibnamefont
  {Kafri}}, \bibinfo {author} {\bibfnamefont {J.~M.}\ \bibnamefont {Taylor}}, \
  and\ \bibinfo {author} {\bibfnamefont {G.~J.}\ \bibnamefont {Milburn}},\
  }\href {\doibase 10.1088/1367-2630/16/6/065020} {\bibfield  {journal}
  {\bibinfo  {journal} {New Journal of Physics}\ }\textbf {\bibinfo {volume}
  {16}},\ \bibinfo {pages} {065020} (\bibinfo {year}
  {2014}{\natexlab{a}})}\BibitemShut {NoStop}%
\bibitem [{\citenamefont {Tilloy}\ and\ \citenamefont
  {Di{\'o}si}(2017)}]{tilloyPRD2017}%
  \BibitemOpen
  \bibfield  {author} {\bibinfo {author} {\bibfnamefont {A.}~\bibnamefont
  {Tilloy}}\ and\ \bibinfo {author} {\bibfnamefont {L.}~\bibnamefont
  {Di{\'o}si}},\ }\href {\doibase 10.1103/PhysRevD.96.104045} {\bibfield
  {journal} {\bibinfo  {journal} {Physical Review D}\ }\textbf {\bibinfo
  {volume} {96}},\ \bibinfo {pages} {104045} (\bibinfo {year}
  {2017})}\BibitemShut {NoStop}%
\bibitem [{\citenamefont {Altamirano}\ \emph {et~al.}(2017)\citenamefont
  {Altamirano}, \citenamefont {Corona-Ugalde}, \citenamefont {Mann},\ and\
  \citenamefont {Zych}}]{altamirano2017unitarity}%
  \BibitemOpen
  \bibfield  {author} {\bibinfo {author} {\bibfnamefont {N.}~\bibnamefont
  {Altamirano}}, \bibinfo {author} {\bibfnamefont {P.}~\bibnamefont
  {Corona-Ugalde}}, \bibinfo {author} {\bibfnamefont {R.~B.}\ \bibnamefont
  {Mann}}, \ and\ \bibinfo {author} {\bibfnamefont {M.}~\bibnamefont {Zych}},\
  }\href@noop {} {\bibfield  {journal} {\bibinfo  {journal} {New Journal of
  Physics}\ }\textbf {\bibinfo {volume} {19}},\ \bibinfo {pages} {013035}
  (\bibinfo {year} {2017})}\BibitemShut {NoStop}%
\bibitem [{\citenamefont {Gaona-Reyes}\ \emph {et~al.}(2021)\citenamefont
  {Gaona-Reyes}, \citenamefont {Carlesso},\ and\ \citenamefont
  {Bassi}}]{BassiKTM21}%
  \BibitemOpen
  \bibfield  {author} {\bibinfo {author} {\bibfnamefont {J.~L.}\ \bibnamefont
  {Gaona-Reyes}}, \bibinfo {author} {\bibfnamefont {M.}~\bibnamefont
  {Carlesso}}, \ and\ \bibinfo {author} {\bibfnamefont {A.}~\bibnamefont
  {Bassi}},\ }\href {\doibase 10.1103/PhysRevD.103.056011} {\bibfield
  {journal} {\bibinfo  {journal} {Phys. Rev. D}\ }\textbf {\bibinfo {volume}
  {103}},\ \bibinfo {pages} {056011} (\bibinfo {year} {2021})}\BibitemShut
  {NoStop}%
\bibitem [{\citenamefont {Di~Bartolomeo}\ \emph {et~al.}(2021)\citenamefont
  {Di~Bartolomeo}, \citenamefont {Carlesso},\ and\ \citenamefont
  {Bassi}}]{DissKTM}%
  \BibitemOpen
  \bibfield  {author} {\bibinfo {author} {\bibfnamefont {G.}~\bibnamefont
  {Di~Bartolomeo}}, \bibinfo {author} {\bibfnamefont {M.}~\bibnamefont
  {Carlesso}}, \ and\ \bibinfo {author} {\bibfnamefont {A.}~\bibnamefont
  {Bassi}},\ }\href {\doibase 10.1103/PhysRevD.104.104027} {\bibfield
  {journal} {\bibinfo  {journal} {Phys. Rev. D}\ }\textbf {\bibinfo {volume}
  {104}},\ \bibinfo {pages} {104027} (\bibinfo {year} {2021})}\BibitemShut
  {NoStop}%
\bibitem [{\citenamefont {Bassi}\ \emph {et~al.}(2017)\citenamefont {Bassi},
  \citenamefont {Gro{\ss}ardt},\ and\ \citenamefont {Ulbricht}}]{bassiCQG2017}%
  \BibitemOpen
  \bibfield  {author} {\bibinfo {author} {\bibfnamefont {A.}~\bibnamefont
  {Bassi}}, \bibinfo {author} {\bibfnamefont {A.}~\bibnamefont {Gro{\ss}ardt}},
  \ and\ \bibinfo {author} {\bibfnamefont {H.}~\bibnamefont {Ulbricht}},\
  }\href {\doibase 10.1088/1361-6382/aa864f} {\bibfield  {journal} {\bibinfo
  {journal} {Classical and Quantum Gravity}\ }\textbf {\bibinfo {volume}
  {34}},\ \bibinfo {pages} {193002} (\bibinfo {year} {2017})}\BibitemShut
  {NoStop}%
\bibitem [{\citenamefont {Nielsen}\ and\ \citenamefont
  {Chuang}(2010)}]{nielsen2010}%
  \BibitemOpen
  \bibfield  {author} {\bibinfo {author} {\bibfnamefont {M.~A.}\ \bibnamefont
  {Nielsen}}\ and\ \bibinfo {author} {\bibfnamefont {I.~L.}\ \bibnamefont
  {Chuang}},\ }\href@noop {} {\emph {\bibinfo {title} {Quantum {Computation}
  and {Quantum} {Information}: 10th {Anniversary} {Edition}}}},\ \bibinfo
  {edition} {anniversary edizione}\ ed.\ (\bibinfo  {publisher} {Cambridge
  University Press},\ \bibinfo {address} {Cambridge ; New York},\ \bibinfo
  {year} {2010})\BibitemShut {NoStop}%
\bibitem [{\citenamefont {Ollivier}\ and\ \citenamefont
  {Zurek}(2001)}]{ollivierPRL2001}%
  \BibitemOpen
  \bibfield  {author} {\bibinfo {author} {\bibfnamefont {H.}~\bibnamefont
  {Ollivier}}\ and\ \bibinfo {author} {\bibfnamefont {W.~H.}\ \bibnamefont
  {Zurek}},\ }\href {\doibase 10.1103/PhysRevLett.88.017901} {\bibfield
  {journal} {\bibinfo  {journal} {Phys. Rev. Lett.}\ }\textbf {\bibinfo
  {volume} {88}},\ \bibinfo {pages} {017901} (\bibinfo {year}
  {2001})}\BibitemShut {NoStop}%
\bibitem [{\citenamefont {Henderson}\ and\ \citenamefont
  {Vedral}(2001)}]{hendersonJPAMG2001}%
  \BibitemOpen
  \bibfield  {author} {\bibinfo {author} {\bibfnamefont {L.}~\bibnamefont
  {Henderson}}\ and\ \bibinfo {author} {\bibfnamefont {V.}~\bibnamefont
  {Vedral}},\ }\href {\doibase 10.1088/0305-4470/34/35/315} {\bibfield
  {journal} {\bibinfo  {journal} {J. Phys. A: Math. Gen.}\ }\textbf {\bibinfo
  {volume} {34}},\ \bibinfo {pages} {6899} (\bibinfo {year}
  {2001})}\BibitemShut {NoStop}%
\bibitem [{\citenamefont {Modi}\ \emph {et~al.}(2012)\citenamefont {Modi},
  \citenamefont {Brodutch}, \citenamefont {Cable}, \citenamefont {Paterek},\
  and\ \citenamefont {Vedral}}]{modiRMP2012}%
  \BibitemOpen
  \bibfield  {author} {\bibinfo {author} {\bibfnamefont {K.}~\bibnamefont
  {Modi}}, \bibinfo {author} {\bibfnamefont {A.}~\bibnamefont {Brodutch}},
  \bibinfo {author} {\bibfnamefont {H.}~\bibnamefont {Cable}}, \bibinfo
  {author} {\bibfnamefont {T.}~\bibnamefont {Paterek}}, \ and\ \bibinfo
  {author} {\bibfnamefont {V.}~\bibnamefont {Vedral}},\ }\href {\doibase
  10.1103/RevModPhys.84.1655} {\bibfield  {journal} {\bibinfo  {journal} {Rev.
  Mod. Phys.}\ }\textbf {\bibinfo {volume} {84}},\ \bibinfo {pages} {1655}
  (\bibinfo {year} {2012})}\BibitemShut {NoStop}%
\bibitem [{\citenamefont {Bera}\ \emph {et~al.}(2017)\citenamefont {Bera},
  \citenamefont {Das}, \citenamefont {Sadhukhan}, \citenamefont {Roy},
  \citenamefont {Sen(De)},\ and\ \citenamefont {Sen}}]{bera2017quantum}%
  \BibitemOpen
  \bibfield  {author} {\bibinfo {author} {\bibfnamefont {A.}~\bibnamefont
  {Bera}}, \bibinfo {author} {\bibfnamefont {T.}~\bibnamefont {Das}}, \bibinfo
  {author} {\bibfnamefont {D.}~\bibnamefont {Sadhukhan}}, \bibinfo {author}
  {\bibfnamefont {S.~S.}\ \bibnamefont {Roy}}, \bibinfo {author} {\bibfnamefont
  {A.}~\bibnamefont {Sen(De)}}, \ and\ \bibinfo {author} {\bibfnamefont
  {U.}~\bibnamefont {Sen}},\ }\href {\doibase 10.1088/1361-6633/aa872f}
  {\bibfield  {journal} {\bibinfo  {journal} {Reports on Progress in Physics}\
  }\textbf {\bibinfo {volume} {81}},\ \bibinfo {pages} {024001} (\bibinfo
  {year} {2017})}\BibitemShut {NoStop}%
\bibitem [{\citenamefont {Ciccarello}\ and\ \citenamefont
  {Giovannetti}(2012)}]{ciccarelloPRA2012a}%
  \BibitemOpen
  \bibfield  {author} {\bibinfo {author} {\bibfnamefont {F.}~\bibnamefont
  {Ciccarello}}\ and\ \bibinfo {author} {\bibfnamefont {V.}~\bibnamefont
  {Giovannetti}},\ }\href {\doibase 10.1103/PhysRevA.85.010102} {\bibfield
  {journal} {\bibinfo  {journal} {Phys. Rev. A}\ }\textbf {\bibinfo {volume}
  {85}},\ \bibinfo {pages} {010102} (\bibinfo {year} {2012})}\BibitemShut
  {NoStop}%
\bibitem [{\citenamefont {Streltsov}\ \emph {et~al.}(2011)\citenamefont
  {Streltsov}, \citenamefont {Kampermann},\ and\ \citenamefont
  {Bruß}}]{streltsovPRL2011}%
  \BibitemOpen
  \bibfield  {author} {\bibinfo {author} {\bibfnamefont {A.}~\bibnamefont
  {Streltsov}}, \bibinfo {author} {\bibfnamefont {H.}~\bibnamefont
  {Kampermann}}, \ and\ \bibinfo {author} {\bibfnamefont {D.}~\bibnamefont
  {Bruß}},\ }\href {\doibase 10.1103/PhysRevLett.107.170502} {\bibfield
  {journal} {\bibinfo  {journal} {Phys. Rev. Lett.}\ }\textbf {\bibinfo
  {volume} {107}},\ \bibinfo {pages} {170502} (\bibinfo {year}
  {2011})}\BibitemShut {NoStop}%
\bibitem [{\citenamefont {Giorda}\ and\ \citenamefont
  {Paris}(2010)}]{giordaPRL2010}%
  \BibitemOpen
  \bibfield  {author} {\bibinfo {author} {\bibfnamefont {P.}~\bibnamefont
  {Giorda}}\ and\ \bibinfo {author} {\bibfnamefont {M.~G.~A.}\ \bibnamefont
  {Paris}},\ }\href {\doibase 10.1103/PhysRevLett.105.020503} {\bibfield
  {journal} {\bibinfo  {journal} {Phys. Rev. Lett.}\ }\textbf {\bibinfo
  {volume} {105}},\ \bibinfo {pages} {020503} (\bibinfo {year}
  {2010})}\BibitemShut {NoStop}%
\bibitem [{\citenamefont {Adesso}\ and\ \citenamefont
  {Datta}(2010)}]{adessoPRL2010}%
  \BibitemOpen
  \bibfield  {author} {\bibinfo {author} {\bibfnamefont {G.}~\bibnamefont
  {Adesso}}\ and\ \bibinfo {author} {\bibfnamefont {A.}~\bibnamefont {Datta}},\
  }\href {\doibase 10.1103/PhysRevLett.105.030501} {\bibfield  {journal}
  {\bibinfo  {journal} {Phys. Rev. Lett.}\ }\textbf {\bibinfo {volume} {105}},\
  \bibinfo {pages} {030501} (\bibinfo {year} {2010})}\BibitemShut {NoStop}%
\bibitem [{\citenamefont {Kafri}\ \emph
  {et~al.}(2014{\natexlab{b}})\citenamefont {Kafri}, \citenamefont {Taylor},\
  and\ \citenamefont {Milburn}}]{kafriNJP2014a}%
  \BibitemOpen
  \bibfield  {author} {\bibinfo {author} {\bibfnamefont {D.}~\bibnamefont
  {Kafri}}, \bibinfo {author} {\bibfnamefont {J.~M.}\ \bibnamefont {Taylor}}, \
  and\ \bibinfo {author} {\bibfnamefont {G.~J.}\ \bibnamefont {Milburn}},\
  }\href {\doibase 10.1088/1367-2630/16/6/065020} {\bibfield  {journal}
  {\bibinfo  {journal} {New Journal of Physics}\ }\textbf {\bibinfo {volume}
  {16}},\ \bibinfo {pages} {065020} (\bibinfo {year}
  {2014}{\natexlab{b}})}\BibitemShut {NoStop}%
\bibitem [{\citenamefont {Altamirano}\ \emph {et~al.}(2018)\citenamefont
  {Altamirano}, \citenamefont {Corona-Ugalde}, \citenamefont {Mann},\ and\
  \citenamefont {Zych}}]{Altamirano_2018}%
  \BibitemOpen
  \bibfield  {author} {\bibinfo {author} {\bibfnamefont {N.}~\bibnamefont
  {Altamirano}}, \bibinfo {author} {\bibfnamefont {P.}~\bibnamefont
  {Corona-Ugalde}}, \bibinfo {author} {\bibfnamefont {R.~B.}\ \bibnamefont
  {Mann}}, \ and\ \bibinfo {author} {\bibfnamefont {M.}~\bibnamefont {Zych}},\
  }\href {\doibase 10.1088/1361-6382/aac72f} {\bibfield  {journal} {\bibinfo
  {journal} {Classical and Quantum Gravity}\ }\textbf {\bibinfo {volume}
  {35}},\ \bibinfo {pages} {145005} (\bibinfo {year} {2018})}\BibitemShut
  {NoStop}%
\bibitem [{\citenamefont {Krisnanda}\ \emph {et~al.}(2020)\citenamefont
  {Krisnanda}, \citenamefont {Tham}, \citenamefont {Paternostro},\ and\
  \citenamefont {Paterek}}]{krisnandaNQI2020}%
  \BibitemOpen
  \bibfield  {author} {\bibinfo {author} {\bibfnamefont {T.}~\bibnamefont
  {Krisnanda}}, \bibinfo {author} {\bibfnamefont {G.~Y.}\ \bibnamefont {Tham}},
  \bibinfo {author} {\bibfnamefont {M.}~\bibnamefont {Paternostro}}, \ and\
  \bibinfo {author} {\bibfnamefont {T.}~\bibnamefont {Paterek}},\ }\href
  {\doibase 10.1038/s41534-020-0243-y} {\bibfield  {journal} {\bibinfo
  {journal} {npj Quantum Information}\ }\textbf {\bibinfo {volume} {6}},\
  \bibinfo {pages} {1} (\bibinfo {year} {2020})}\BibitemShut {NoStop}%
\bibitem [{\citenamefont {Ferraro}\ \emph {et~al.}(2005)\citenamefont
  {Ferraro}, \citenamefont {Olivares},\ and\ \citenamefont
  {Paris}}]{ferraro2005}%
  \BibitemOpen
  \bibfield  {author} {\bibinfo {author} {\bibfnamefont {A.}~\bibnamefont
  {Ferraro}}, \bibinfo {author} {\bibfnamefont {S.}~\bibnamefont {Olivares}}, \
  and\ \bibinfo {author} {\bibfnamefont {M.~G.~A.}\ \bibnamefont {Paris}},\
  }\href {https://air.unimi.it/handle/2434/15713#.XRDX9y2B1QI} {\emph {\bibinfo
  {title} {Gaussian states in quantum information}}}\ (\bibinfo  {publisher}
  {Bibliopolis},\ \bibinfo {year} {2005})\BibitemShut {NoStop}%
\bibitem [{\citenamefont {Purkayastha}(2022)}]{purkayastha2022lyapunov}%
  \BibitemOpen
  \bibfield  {author} {\bibinfo {author} {\bibfnamefont {A.}~\bibnamefont
  {Purkayastha}},\ }\href@noop {} {\bibfield  {journal} {\bibinfo  {journal}
  {arXiv preprint arXiv:2201.00677}\ } (\bibinfo {year} {2022})}\BibitemShut
  {NoStop}%
\bibitem [{\citenamefont {Cover}\ and\ \citenamefont
  {Thomas}(2006)}]{cover2006}%
  \BibitemOpen
  \bibfield  {author} {\bibinfo {author} {\bibfnamefont {T.~M.}\ \bibnamefont
  {Cover}}\ and\ \bibinfo {author} {\bibfnamefont {J.~A.}\ \bibnamefont
  {Thomas}},\ }\href@noop {} {\emph {\bibinfo {title} {Elements of
  {Information} {Theory}}}},\ \bibinfo {edition} {2nd}\ ed.\ (\bibinfo
  {publisher} {Wiley-Interscience},\ \bibinfo {address} {Hoboken, N.J},\
  \bibinfo {year} {2006})\BibitemShut {NoStop}%
\bibitem [{\citenamefont {Pirandola}\ \emph {et~al.}(2014)\citenamefont
  {Pirandola}, \citenamefont {Spedalieri}, \citenamefont {Braunstein},
  \citenamefont {Cerf},\ and\ \citenamefont {Lloyd}}]{pirandolaPRL2014}%
  \BibitemOpen
  \bibfield  {author} {\bibinfo {author} {\bibfnamefont {S.}~\bibnamefont
  {Pirandola}}, \bibinfo {author} {\bibfnamefont {G.}~\bibnamefont
  {Spedalieri}}, \bibinfo {author} {\bibfnamefont {S.~L.}\ \bibnamefont
  {Braunstein}}, \bibinfo {author} {\bibfnamefont {N.~J.}\ \bibnamefont
  {Cerf}}, \ and\ \bibinfo {author} {\bibfnamefont {S.}~\bibnamefont {Lloyd}},\
  }\href {\doibase 10.1103/PhysRevLett.113.140405} {\bibfield  {journal}
  {\bibinfo  {journal} {Phys. Rev. Lett.}\ }\textbf {\bibinfo {volume} {113}},\
  \bibinfo {pages} {140405} (\bibinfo {year} {2014})}\BibitemShut {NoStop}%
\bibitem [{\citenamefont {Gerry}\ \emph {et~al.}(2005)\citenamefont {Gerry},
  \citenamefont {Knight},\ and\ \citenamefont
  {Knight}}]{gerry2005introductory}%
  \BibitemOpen
  \bibfield  {author} {\bibinfo {author} {\bibfnamefont {C.}~\bibnamefont
  {Gerry}}, \bibinfo {author} {\bibfnamefont {P.}~\bibnamefont {Knight}}, \
  and\ \bibinfo {author} {\bibfnamefont {P.~L.}\ \bibnamefont {Knight}},\
  }\href@noop {} {\emph {\bibinfo {title} {Introductory quantum optics}}}\
  (\bibinfo  {publisher} {Cambridge university press},\ \bibinfo {year}
  {2005})\BibitemShut {NoStop}%
\end{thebibliography}%
		\bibliographystyle{apsrev4-1}

\end{document}